\documentclass[aps,prl,twocolumn,showpacs,superscriptaddress,floatfix,10pt]{revtex4-1}

\usepackage{graphicx}
\usepackage{amsmath}
\usepackage{epsfig}
\usepackage{helvet}
\usepackage{amssymb}
\usepackage{epstopdf}

\newcommand{\bra}[1]{\mbox{$\langle #1 |$}}
\newcommand{\ket}[1]{\mbox{$| #1 \rangle$}}

\newcommand{\norm}[1]{\mbox{$\left\| #1 \right\|$}}
\newcommand{\frw}[1]{$\overset{\lower0.5em\hbox{$\smash{\scriptscriptstyle\smile}$}} #1$}

\begin{document}

\title{Entanglement renormalization and wavelets}

\author{Glen Evenbly}
\author{Steven R. White}
\affiliation{Department of Physics and Astronomy, University of California, Irvine, CA 92697-4575 USA}
\date{\today}

\begin{abstract}
We establish a precise connection between discrete wavelet transforms (WTs) and entanglement renormalization (ER), a real-space renormalization group
transformation for quantum systems on the lattice, in the context of free
particle systems. Specifically, we employ Daubechies wavelets to build
approximations to the ground state of the critical Ising model, then
demonstrate that these states correspond to instances of the multi-scale
entanglement renormalization ansatz (MERA), producing the first known analytic MERA for critical systems.  
\end{abstract}

\maketitle
In recent years tensor networks \cite{TNS} have emerged as an exciting
approach to both quantum mechanics and statistical mechanics that
combine ideas of many-body physics with quantum information, while
closely connecting simulation and analytic theory. An intriguing development
within tensor networks is the multi-scale entanglement renormalization ansatz
(MERA) \cite{ER,MERA}, designed to implement real-space renormalization group
(RG) \cite{RGreview} ideas in a powerful numerical algorithm which accurately
captures scale invariance and critical point behavior.

For a $D$-dimensional physical system, the MERA is constructed as a $(D+1)$-dimensional
tensor network, where layers in the extra dimension encode ground state correlations at different length scales. Within 
a numerical setting, MERA have been 
demonstrated \cite{MERAlocal, MERAnonlocal, MERAbook, Alg1, Alg2, MERAapp1, MERAapp2, MERAapp3} to 
accurately capture the critical long range behavior of lattice versions of
conformal field theories (CFTs) \cite{CFTbook1,CFTbook2}, 
which are used to describe critical points. MERA also provides a framework to investigate the AdS/CFT correspondence, with the
extra dimension of the MERA associated with a physical space-time
dimension, making tensor networks an important topic in quantum
gravity and string theory \cite{Holo1,Holo2}.

Wavelets and wavelet transforms (WTs) \cite{Daub1,Daub2,Daub3,Daub4,WaveBook}, 
one of the most significant
developments in signal and image processing in several decades,
are also closely tied to RG: ideas from RG influenced the development of wavelets, and
wavelets have proved to be a useful tool in RG applications \cite{Battle}.
In fact, it is natural to think of compact, orthogonal WTs, such as the well-known families
of WTs introduced by Daubechies\cite{Daub1,Daub4}, as being real-space RG transformations, but
in the space of ordinary 1D functions rather than in terms of
Hamiltonians or Lagrangians.
Given the close connections to real-space RG of both MERA and wavelets,  
it is natural to ask if these two methods are connected more deeply to each other.

Here we show that this is, indeed, the case, and report on a precise relation
between WTs and the MERA; that a wavelet analysis of a free particle system can
be exactly mapped to a MERA.  An important development results
from this connection:  we find the first {\it analytic}  MERA for
critical systems, whereas previous constructions have
always resulted from a complicated variational optimization. 
For the lowest order case, the isometries $w$ and disentanglers $u$ that constitute a scale-invariant MERA can be written in a remarkably compact form,
\begin{align}
  w & = \tfrac{{\sqrt 3  + \sqrt 2 }}{4}II + \tfrac{{\sqrt 3  - \sqrt 2 }}{4}ZZ + \tfrac{{i(1 + \sqrt 2 )}}{4}XY + \tfrac{{i(1 - \sqrt 2 )}}{4}YX \hfill \nonumber\\
  u & = \tfrac{{\sqrt 3  + 2}}{4}II + \tfrac{{\sqrt 3  - 2}}{4}ZZ + \tfrac{i}{4}XY + \tfrac{i}{4}YX, \label{eq:16}
\end{align}
where $X$, $Y$, $Z$ are Pauli matrices, and
where $ZZ$ is short for $Z\otimes Z$ etc. 
This MERA, which is derived using two copies of the Daubechies D4 wavelet, 
can be shown to approximate the ground state of the quantum critical Ising model, 
including the critical data of the Ising CFT. 
We also present a general prescription for obtaining higher order MERA and give another specific example.
These analytic constructions provide an important
new tool for making further progress in MERA applications, and are
likely just the first example of important results from the
wavelet-MERA connection.

\textit{Free Fermions.---} 
We consider the tight-binding Hamiltonian on an infinite $1D$ lattice of spinless fermions at half-filling,
\begin{equation}
H_\textrm{ff} = -\sum\nolimits_r {\left( {\hat a_{r + 1}^\dag {{\hat a}_r} + \hat a_r^\dag {{\hat a}_{r + 1}}} \right)}, \label{eq:1}
\end{equation}
with $\hat a$ and $\hat a^\dag$ the annihilation and creation operators. 
In terms of the Fourier mode creation operators $\hat b_k ^\dag$, where $k$ is the momentum, the ground state of $H$ is given by occupying the negative energy modes, 
\begin{equation}
\left| {{\psi _\textrm{GS}}} \right\rangle  =  {\prod\limits_{\left| k \right| < \pi /2} {{{\hat b}_k^\dag}} } \left| 0 \right\rangle \label{eq:4}
\end{equation}
while leaving positive energy modes unoccupied. 

\begin{figure}[!t]
\begin{center}
\includegraphics[width=8cm]{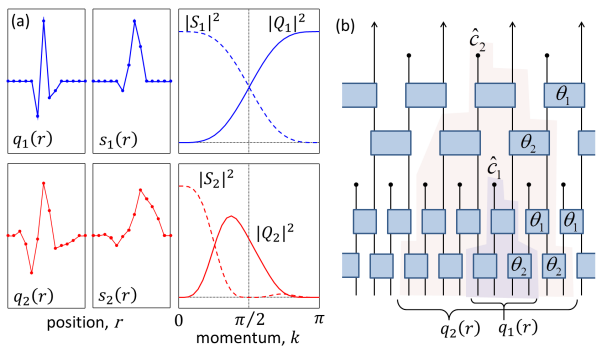}
\caption{(a) Plots of D4 Daubechies wavelets $q_z(r)$ and scaling functions
$s_z(r)$ for scales $z = 1,2$, together with their Fourier spectra $Q_z(k)$
and $S_z(k)$. (b) The quantum circuit, built from gates $v(\theta)$ of 
Eq.  \ref{eq:13} with angles $\theta_1=\pi/12$ and $\theta_2=-\pi/6$, implements the
linear map of fermionic modes, see Eq. \ref{eq:15}, corresponding to the D4
wavelet transform.} \label{fig:BinCircuit}
\end{center}
\end{figure}

\textit{Wavelets.---} We would like to use wavelets to find another set of 
modes $\{ \hat c_z^\textrm{Low} \}$, 
more localized than plane waves, but composed almost entirely of linear 
combinations of negative energy states, so that the ground state is approximated by filling these modes,
\begin{equation}
\left| {{\psi _\textrm{GS}}} \right\rangle  \approx  {\prod\limits_{z} { \left({{\hat c}_z^\textrm{Low}}\right)^\dag } } \left| 0 \right\rangle \label{eq:12}
\end{equation}
Correspondingly, there will also exist a set of modes $\{ \hat c_z^\textrm{High}
\}$ composed of linear combinations of positive energy states, which are
unoccupied in the ground state. The accuracy of the separation into low and
high energy states must be increasingly sharp as one looks more closely near
the Fermi surface, which suggests we require a WT that targets the Fermi
points, $\pm \pi/2$. On the other hand, standard WTs, such as Daubechies
wavelets \cite{Daub1, Daub2}, are designed to approximately divide the Fourier
components into low (scaling function) and high (wavelet) parts at each scale,
see Fig. \ref{fig:BinCircuit}(a), such that they resolve degrees of freedom
close to momentum $k=0$. Thus a direct application of known WTs is not
sufficient to approximate the ground state $\left| {{\psi _\textrm{GS}}}
\right\rangle$. However, we will show that it is possible to combine two
slightly modified Daubechies WTs to give an excellent separation of negative
and positive energies, targeting $k = \pm \pi/2$.

Let us consider the Daubechies D4 wavelets. 
We denote by $q_z(r)$ the wavelet function at scale $z$, and $Q_z(k)$ its (discrete) 
Fourier transform. At the smallest $z=1$ scale, the D4 wavelet
has support on 4 sites, where it has values $q_1 \approx
[-0.483, 0.837, -0.224, -0.129]$, while wavelets at larger scales $q_z$, which
are supported on intervals of $r = 2^{z+1}+2^z-2$ sites, can be easily obtained
from $q_1$ using the cascade algorithm \cite{WaveBook}. At level $z$ the basis includes 
all translations
of $q_z$ by $d = n 2^z$ sites (for integer $n$). 
The set of all wavelets--every level $z$, with all appropriate translations, form a complete,
orthonormal basis of functions. The Daubechies wavelets are designed as high-pass filters, 
such that they are orthogonal to smooth functions (or, equivalently, functions that only 
possess low frequency components). 
Specifically, the D4 wavelets have two vanishing moments about $k = 0$,
\begin{equation}
{Q_z}\left( 0 \right) = 0, \;\;\;\;\;\; {\left. {\frac{{{\partial}
{Q_z}}}{{\partial {k }}}} \right|_{k  = 0}} = 0,  \label{eq:8}
\end{equation}
for all scales $z$, see also Fig. \ref{fig:BinCircuit}(a).

We now construct modified wavelets which have the vanishing moments 
at the Fermi points, $k \pm \pi/2$, rather than $k=0$.
We first we multiply the wavelets by
a phase $\omega(r)=\exp(i\pi r)$ and then dilate by a factor of two, but with the 
in-between sites set to zero(!):
\begin{equation}
\tilde q_z^{{\text{odd}}}(r) = \left\{ {\begin{array}{*{20}{l}}
  {{{\left( { - 1} \right)}^{\left( {\tfrac{{r + 1}}{2}} \right)}}{q_z}\left( {\tfrac{{r + 1}}{2}} \right),}&{r\;\; \textrm{odd}} \\ 
  {0,}&{r\;\; \textrm{even}} 
\end{array}} \right. \label{eq:9}
\end{equation}
Similarly we construct wavelets $\tilde q_z^{{\text{even}}}$ that only 
have support on the even sublattice. 
This transformation into odd and even sublattice parts seems more natural if one considers the 
real linear combinations of the Fermi points $\sin (r \pi/2)$ and $\cos (r \pi/2)$, which are 
zero on even and odd sublattices, respectively.
The key result is that the 
frequency space representation of these 
wavelets $\tilde Q_z^{{\text{odd}}}(k)$ now have vanishing moments at $k = \pm \pi/2$,
\begin{equation}
{{\tilde Q}_z^\textrm{odd}}\left( { \pm \frac{\pi }{2}} \right) = 0,\quad {\left. {\frac{{{\partial  }{{\tilde Q}_z^\textrm{odd}}}}{{\partial {k  }}}} \right|_{ \pm \tfrac{\pi}{2}}} = 0, \label{eq:10}
\end{equation}
and similarly for $\tilde Q_z^{{\text{even}}}$. 

The modified wavelet functions are localized, in Fourier space, increasingly
close to the Fermi points at larger scale $z$; however, they are still not
sufficient to approximate the ground state 
$\left| {{\psi _\textrm{GS}}} \right\rangle$ as they contain a mixture of negative and
positive energy components. To separate the energy components, we form coherent
low $l_z(r)$ and high ${h_z}(r)$ wavelet pairs by taking 
symmetric and anti-symmetric combinations respectively of $\tilde q_z^{{\text{odd}}}$ and $\tilde q_z^{{\text{even}}}$,
\begin{align}
  {l_z}(r) &= {{\tilde q}_z^\textrm{odd}}(r) + {{\tilde q^\textrm{even}}_z}({r_0} - r) \hfill \nonumber \\
  {h_z}(r) &= {{\tilde q}_z^\textrm{odd}}(r) - {{\tilde q^\textrm{even}}_z}({r_0} - r) \hfill. \label{eq:11}
\end{align}
The constant $r_0$, which determines the spatial alignment of the odd and even
wavelets, should be chosen in order to form wavelets with the best separation of energies; here this choice is such that the
support of an even wavelet starts three sites before that of an odd wavelet it is paired with (see Ref.\cite{Supp} Section A for details).

\begin{figure}[!t]
\begin{center}
\includegraphics[width=8cm]{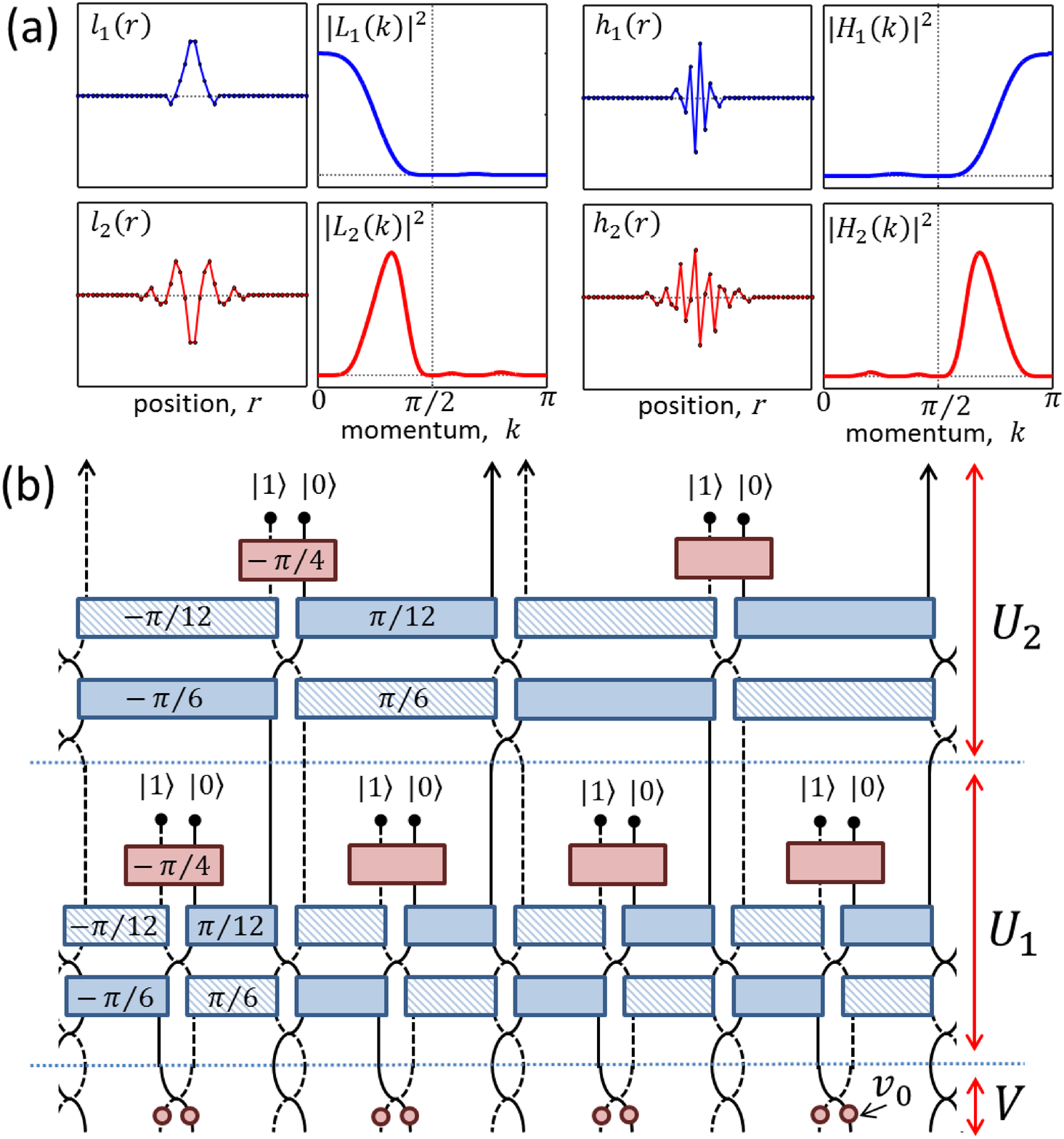}
\caption{(a) Plots of low frequency ${l_z}(r)$ and high frequency ${h_z}(r)$
wavelets from Eq. \ref{eq:11}, together with their Fourier spectra $L_z(k)$
and $H_z(k)$, for scales $z = 1,2$. (b) Quantum circuit that approximates
the free fermion ground state by setting modes corresponding to low frequency
${l_z}$ wavelets in the occupied $\ket{1}$ state and high frequency ${h_z}$
wavelets in the unoccupied $\ket{0}$ state. The circuit is built from gates
$v(\theta)$ as defined in Eq.  \ref{eq:13} with angles $\theta$ as indicated,
and $v_0$ represents a phase gate with angle $\pi$.} 
\label{fig:DoubleCircuit}
\end{center}
\end{figure}

The symmetric and anti-symmetric wavelet pairs ${l_z}(r)$ and ${h_z}(r)$,
together with their frequency spectra, are plotted in
Fig. \ref{fig:DoubleCircuit}(a) for scales $z=1,2$. It can be seen that they
separate negative and positive energies nicely, with ${l_z}(r)$, to very good approximation, 
only containing 
frequencies $|k|<\pi/2$ (and vice-versa for ${h_z}(r)$). 
Thus, if we use the wavelets ${l_z}(r)$ and ${h_z}(r)$ to define a linear mapping of fermionic modes,
\begin{align}
\hat c_z^\textrm{Low} &= \sum\limits_r {{{\hat a}_r}{l_z}(r)} ,\nonumber \\  
\hat c_z^\textrm{High} &= \sum\limits_r {{{\hat a}_r}{h_z}(r)}, \label{eq:5}
\end{align}
then the ground state $\left| {{\psi _\textrm{GS}}} \right\rangle$
of the free fermion model $H_\textrm{ff}$ can be approximated by occupying the
negative energy modes $\hat c_z^\textrm{Low} $ as per Eq. \ref{eq:12}.

\textit{Quantum circuit.---} Through Eqs. \ref{eq:9}, \ref{eq:11} and \ref{eq:5} we have identified a discrete wavelet transform of fermionic modes that can be used to approximate the free fermion ground state. We now describe how this transform of fermionic modes can be realized as a quantum circuit, employing a formalism similar to that of Refs.\cite{MERAferm, FFTandMERA, MattWavelet}, and argue that this circuit corresponds precisely to a MERA. 

We restrict to a circuit built from two-site unitary gates $v$ that preserve
particle number,
\begin{equation}
{v_{r,r + 1}}\left( \theta  \right) = \left[ {\begin{array}{*{20}{c}}
  1&0&0&0 \\ 
  0&{\cos \left( \theta  \right)}&{ - \sin \left( \theta  \right)}&0 \\ 
  0&{\sin \left( \theta  \right)}&{\cos \left( \theta  \right)}&0 \\ 
  0&0&0&1 
\end{array}} \right] \label{eq:13}
\end{equation}
written in the number basis 
$\left\{ {\left| {00} \right\rangle ,\left| {01} \right\rangle
,\left| {10} \right\rangle ,\left| {11} \right\rangle } \right\}$ 
for some angle $\theta \in [-\pi,\pi]$.  It is known that unitary gates $v(\theta)$ map fermionic modes \emph{linearly}, such that, under action of
$v(\theta)$, a pair of fermionic modes $\hat a_r$ and $\hat a_{r+1}$ is mapped to a
new set of modes $\hat d_r$ and $\hat d_{r+1}$, 
\begin{equation}
\left[ {\begin{array}{*{20}{c}}
  {{{\hat d}_r}} \\ 
  {{{\hat d}_{r + 1}}} 
\end{array}} \right] \equiv \left[ {\begin{array}{*{20}{c}}
  {\cos \left( \theta  \right)}&{\sin \left( \theta  \right)} \\ 
  { - \sin \left( \theta  \right)}&{\cos \left( \theta  \right)} 
\end{array}} \right]\left[ {\begin{array}{*{20}{c}}
  {{{\hat a}_r}} \\ 
  {{{\hat a}_{r + 1}}} 
\end{array}} \right]. \label{eq:14}
\end{equation}
Any unitary linear map on $M$ fermionic modes can be decomposed as a product of such
two-site maps, hence can also be expressed as a quantum circuit built from the
unitary gates $v(\theta)$ of Eq. \ref{eq:13}. It follows that an orthogonal
wavelet transform, which implements a unitary map of fermionic modes
\cite{WaveNorm}, \begin{equation}
{\hat c_z} = \sum\nolimits_r {{{\hat a}_r}{q_z}(r)}, \label{eq:15}
\end{equation}
where ${q_z}(r)$ are the wavelet coefficients at scale $z$, can also be expressed as a unitary circuit built from gates $v(\theta)$. 
Fig. \ref{fig:BinCircuit}(b) shows the circuit diagram for the D4 wavelets, which is built from two distinct unitary gates: $\{v(\pi/12), v(-\pi/6) \}$. The structure of this circuit follows from the fast wavelet transform algorithm \cite{FWT}, which implements the WT on $2^M$ sites through $M$ recursive applications of a filter bank, and thus allows the circuit to be organized into $M$ layers. The filter bank corresponding to a WT of $2N$ coefficients can be further decomposed into a depth $N$ circuit of gates $\{v(\theta_1), v(\theta_2),\ldots, v(\theta_N)\}$, where the angles $\theta_i$ are fixed from the WT under consideration, see Refs.\cite{MattWavelet, LongWavelet} for additional details. Notice that the circuit of 
Fig. \ref{fig:BinCircuit}(b) corresponding to the D4 wavelet is precisely a scale-invariant MERA. In a higher order WT, such as the D2N Daubechies wavelets with $N>2$, the circuit would have $N$ levels of unitary gates $v(\theta_i)$ in each layer and thus no longer correspond to a standard MERA; however under appropriate grouping of gates, see Ref.\cite{Supp} Section B, one could reinterpret this as a MERA of larger bond dimension. 

Two copies of the circuit representation of the D4 wavelets can then be
combined to construct the modified wavelet transform of Eqs. \ref{eq:11} and
\ref{eq:5} that approximates the free fermion ground state, as depicted in
Fig. \ref{fig:DoubleCircuit}(b). Here one copy of the circuit for the D4
WT is implemented on the odd sublattice, and is overlaid with a second
circuit (spatially mirrored with respect to the first) on the even sublattice.
Note that the spatial mirroring is equivalent to negating the sign of the
unitary angles, such that the second circuit is comprised of gates
$\{v(-\pi/12), v(\pi/6) \}$. These two circuits are then coupled by $v(-\pi/4)$
gates, which generate the symmetric/antisymmetric wavelets described in Eq.  \ref{eq:11}.

The combined circuit depicted in Fig. \ref{fig:DoubleCircuit}(b) consists of an identical sequence of scale-invariant layers, labeled $\{U_1,U_2,\ldots \}$, but also includes an initial `transitional' layer $V$ at the bottom. The transitional layer serves two purposes, (i) firstly it includes local unitary operators $v_0 = \hat a^\dag \hat a - \hat a \hat a^\dag$ that implement the phase change described in Eq. \ref{eq:9}, and (ii) secondly, it includes an
extra set of swap gates required to give the proper wavelet alignment $r_0$ in Eq. \ref{eq:11}. Under appropriate grouping of tensors, the circuit of Fig. \ref{fig:DoubleCircuit}(b) can be mapped to a binary MERA of bond dimension $\chi=4$. This MERA can then be split into two copies of the $\chi=2$ MERA from  Eq. \ref{eq:16} for the ground state of the quantum critical Ising model, $H_\textrm{Is.} = \sum\nolimits_r {( - {X_r} {X_{r + 1}} - {Z_r}} )$, using known decoupling \cite{Decouple} and Jordan-Wigner \cite{GroundEnt} transformations, see Ref.\cite{Supp} Section C for details.


\begin{table}[]
\centering
\label{datatable}
\begin{tabular}{c|c|cc}
          & Exact &  MERA & MERA      \\
                    &  &  $\chi = 2$     & $\chi = 8$      \\
          \hline
Energy & \;\;-1.27323\ldots\;\;&\quad \;\;-1.24212\;\; \quad &\quad \;\;-1.26774 \;\; \quad \\    
	   &					& ($2.4\%$ err.)& ($0.4\%$ err.) \\
          \hline
c & 0.5 & 0.4957 & 0.5041 \\ 
\hline
$\Delta_\mathbb{I}$& 0 	  &0      & 0            \\
$\Delta_\sigma$    & 0.125 &0.1402 & 0.1233  \\
$\Delta_\epsilon$  & 1     &1      & 1            \\
$\Delta_\mu$       & 0.125 &0.1445 & 0.1291       \\
$\Delta_\psi$ 	  & 0.5   &0.5    & 0.5    \\
$\Delta_{\bar{\psi}}$& 0.5   &0.5    & 0.5     \\
$\Delta_H$& 2   &2    & 2     \\
\hline
$C_{\epsilon,\sigma,\sigma}$  &   0.5     &   0.4584     &  0.4957    \\
$C_{\epsilon,\mu,\mu}$  &   -0.5    &  -0.4201     &  -0.5060  \\
$C_{\psi,\mu,\sigma}$  &   $\tfrac{{{e^{ - i\pi /4}}}}{{\sqrt 2 }}$    &  $\tfrac{{{1.1422 e^{ - i\pi /4}}}}{{\sqrt 2 }}$    &  $\tfrac{{{1.0014 e^{ - i\pi /4}}}}{{\sqrt 2 }}$ \\
$C_{\bar{\psi},\mu,\sigma}$  &   $\tfrac{{{e^{ i\pi /4}}}}{{\sqrt 2 }}$    &  $\tfrac{{{1.1422 e^{ i\pi /4}}}}{{\sqrt 2 }}$    &  $\tfrac{{{1.0014 e^{ i\pi /4}}}}{{\sqrt 2 }}$ \\
$C_{\epsilon,\psi,\bar{\psi}}$  &   $i$    &  1.234$i$     &  1.0243$i$ \\
$C_{\epsilon,\bar{\psi},\psi}$  &   -$i$    &  -1.234$i$     &  -1.0243$i$   
\end{tabular}
\caption{Energy density, central charge $c$, scaling dimensions $\Delta_i$ of primary fields (and also of the Hamiltonian $\Delta_H$), OPE coefficients $C_{ijk}$, of the $\chi=2,8$ MERA constructed using wavelets.}
\end{table}

\textit{Results and discussion.---} We now analyze the accuracy of the $\chi=2$
MERA from Eq. \ref{eq:16}, constructed using D4 wavelets, and a
$\chi=8$ MERA, constructed using higher order wavelets (see Ref.\cite{Supp},
Section B), as approximate ground states of the quantum critical Ising model.
Note that, as with the $\chi=2$ MERA, the parameters defining the $\chi=8$ MERA
are exactly specified from a closed-form solution. The ground energy and
critical data \cite{CFTbook1, CFTbook2} (including central charge $c$, scaling
dimensions $\Delta_i$ and operator product expansion (OPE) coefficients
$C_{ijk}$) are computed using standard MERA techniques
\cite{MERAlocal,MERAnonlocal,MERAbook}, and the results displayed in Table I.

The wavelet-derived results reproduce the critical data of the Ising CFT, 
with the $\chi=8$ MERA providing significantly better accuracy. However, we find it remarkable that the $\chi=2$ MERA of Eq. \ref{eq:16} does reasonably encode the CFT, despite the simplicity of the tensors it is constructed from. A novel feature of these MERA is that some of the scaling dimensions are reproduced exactly (specifically those corresponding to primary fields 
$\{ \mathbb{I},\varepsilon,\psi,\bar{\psi} \}$, as well as for several of their descendants), which has never been achieved with variationally optimized MERA.

As usual, each layer of the MERA can be understood as implementing a step of entanglement renormalization (ER), which can be used to generate a sequence of increasingly coarse-grained Hamiltonians, 
\begin{equation}
H_\textrm{Is.}^{[0]} \to H_\textrm{Is.}^{[1]} \to H_\textrm{Is.}^{[2]} \to H_\textrm{Is.}^{[3]} \to \ldots, \label{eq:17}
\end{equation}
where $H_\textrm{Is.}^{[z]}$ is the effective Hamiltonian after $z$ steps; see Ref.\cite{Supp} Section D for details. Here we find that the wavelet-derived MERA generates a flow to a gapless fixed point $H_\textrm{Is.}^{*}$ that approximates the thermodynamic limit of the critical Ising model. This is the first known example (analytic
\emph{or} numeric) of a critical Hamiltonian that is coarse-grained to a truly gapless fixed point using ER. In previous (variational) implementations of ER,
the gapless fixed point is approximated for a finite number of RG steps (which
can be increased by using larger $\chi$), before ultimately flowing to a gapped
fixed point \cite{TNR}. This was understood to be an inescapable consequence of
finite bond dimension $\chi$; that truncation errors introduce
relevant perturbations that shift the RG flow off criticality. Here we
have demonstrated that the (previously observed) inability of ER to fully
reproduce a critical RG fixed point stems from a limitation of the optimization
strategies used, as opposed to an inherent limitation of the finite-$\chi$
MERA. This result may hint towards better strategies for numerical optimization of MERA in general. 

The wavelet methodology could be extended to allow analytic construction of MERA (and
potentially branching MERA \cite{Branching}) for free fermions in higher
dimensions, and could also be extended to free bosonic MERA \cite{MERAbose}
which may connect with previous use of wavelets to study bosonic field theories
\cite{FieldWavelets}. We expect that the wavelet-MERA relation will lead to
other useful results, potentially allowing a better characterization of errors
and improved implementations of MERA, and to be useful in the ongoing efforts
to understand MERA in the context of AdS/CFT. Going the other way, this
relation could also lead to useful advances in the design of wavelet transforms and in wavelet applications \cite{LongWavelet}.

The authors acknowledge support by the Simons Foundation (Many Electron
Collaboration). SRW acknowledges funding  from the NSF under grant DMR-1505406.

\newpage

\begin{widetext}
{{\Large {\textbf{ Supplemental Material: Entanglement renormalization and wavelets}}}}
\end{widetext}

\renewcommand\thefigure{A.\arabic{figure}}    
\renewcommand{\theequation}{A.\arabic{equation}}
\setcounter{figure}{0}  
\setcounter{equation}{0} 

\textbf{Section A: Forming coherent low/high energy wavelets.---}
In this section we discuss in more detail a key step in the wavelet
construction for the ground state of free fermions: the formation low/high
energy wavelets by taking symmetric/antisymmetric pairs of Daubechies wavelets.  

\begin{figure}[!b!t!h]
\begin{center}
\includegraphics[width=8cm]{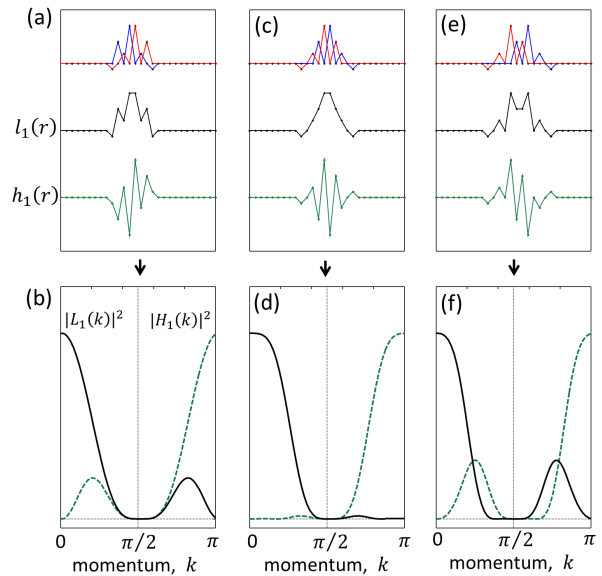}
\caption{(a) Depiction of ${\tilde q}_1^\textrm{odd}$ (blue) and ${\tilde
q}_1^\textrm{even}$ (red), together with symmetric $l_1$ and anti-symmetric
$h_1$ combinations for the alignment $r_0$ depicted, see also Eq. \ref{eq:C1}.
(b) Fourier spectra $L_z$ and $H_z$ of the wavelets $l_z$ and $h_z$ from (a).
The wavelets do not form a coherent low/high energy pair (i.e. $L_z$ still has
significant component for $|k|>\pi/2$, and vice-versa for $H_z$). (c-d) The
wavelets at a slightly different alignment now produce (to good approximation)
a coherent low/high energy pair. This is the alignment used in the derivation
of the ground state wavelet approximation considered in the main text. (e-f) At
a different alignment still, the wavelets again fail to form a coherent
low/high energy pair.} \label{fig:Alignment}
\end{center}
\end{figure}

Given modified D4 Daubechies wavelets supported on the odd ${{\tilde
q}_z^\textrm{odd}}$ and even ${{\tilde q^\textrm{even}}_z}$ sublattices, see
Eq. \ref{eq:9}, we form symmetric $l_z$ and anti-symmetric $h_z$ wavelets as,
\begin{align}
  {l_z}(r) &= {{\tilde q}_z^\textrm{odd}}(r) + {{\tilde q^\textrm{even}}_z}({r_0} - r) \hfill \nonumber \\
  {h_z}(r) &= {{\tilde q}_z^\textrm{odd}}(r) - {{\tilde q^\textrm{even}}_z}({r_0} - r) \hfill, \label{eq:C1}
\end{align}
see also Eq. \ref{eq:11}. Here $r_0$ is a discrete parameter that adjusts the
spatial alignment of the paired wavelets. In practice $r_0$ should be chosen
such that symmetric $l_z$ and anti-symmetric $h_z$ wavelets have the best
separation of energy, i.e. such that $l_z$ only has low frequency components
and $h_z$ only has high frequency components. In Fig. \ref{fig:Alignment} the
wavelets $l_z$ and $h_z$ at scale $z=0$ (together with their corresponding
Fourier spectra $L_z$ and $H_z$) are shown for three different alignments
$r_0$. In all cases the Fourier spectra $L_z$ and $H_z$ vanish at 
$k = \pm \pi/2$, in accordance with Eq. \ref{eq:10}. However, it is only for the proper
alignment, see Fig. \ref{fig:Alignment}(c,d), that the wavelets $l_z$ and $h_z$
form a coherent low/high energy pair to good approximation.

More generally, the best alignment $r_0$ depends on particular the wavelets
$q_z$ in use. In the circuit representation of the wavelet transform, see
Fig. \ref{fig:DoubleCircuit}(b), different alignments can be achieved through
the addition of swap gates in the transitional layer $V$ of the circuit.  

\renewcommand\thefigure{B.\arabic{figure}}    
\renewcommand{\theequation}{B.\arabic{equation}}
\setcounter{figure}{0}  
\setcounter{equation}{0} 

\textbf{Section B: Higher order wavelet approximations.---}
Here we describe a systematic approach for constructing MERA that better
approximate the ground state of free spinless fermions (and the quantum
critical Ising model), based on higher order wavelets. By design, the D4
Daubechies wavelets have two vanishing moments, 
\begin{equation}
{Q_z}\left( 0 \right) = 0, \;\;\;\;\;\; 
{\left. {\frac{{{\partial }{Q_z}}}{{\partial {k }}}} 
\right|_{k  = 0}} = 0,  \label{eq:A1}
\end{equation}
while, more generally, higher order D2N Daubechies wavelets have N vanishing moments.
The D4 wavelet solution of the free fermion model, considered in the main text,
can be easily extended to higher order D2N wavelets by employing the same
method of coupling two copies of the modified wavelet transform, as described
in Eq. \ref{eq:11}). In terms of the circuit representation of the WT, just as
the unitary circuit that implements the D4 wavelets had two levels of unitary
gates $\{v(\theta_1),v(\theta_2)\}$ in each layer, see
Fig. \ref{fig:BinCircuit}(b), a unitary circuit for the D2N Daubechies wavelets can be
built with N levels unitary gates 
$\{v(\theta_1), v(\theta_2), v(\theta_3),\ldots, v(\theta_N) \}$ in each layer 
(see Ref.\cite{LongWavelet}
for details). The circuit with more levels of unitary gates in each layer could
subsequently be mapped to a MERA of larger bond dimension $\chi$, as discussed
later in this section.

\begin{figure}[!t]
\begin{center}
\includegraphics[width=8cm]{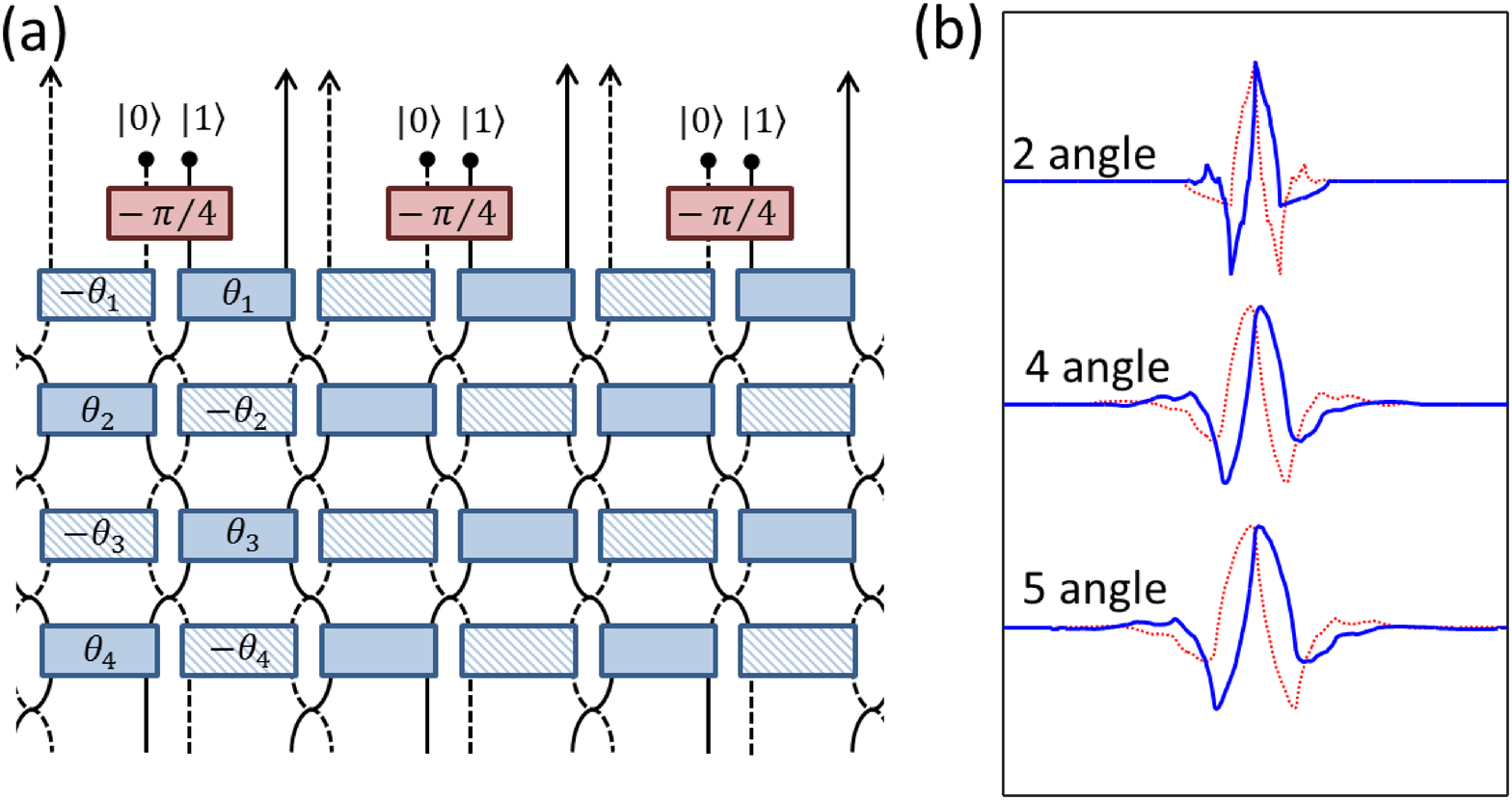}
\caption{(a) One layer of the circuit, which now has four levels of unitary gates, that can encode a higher order wavelet transformation. The full circuit, formed by stacking identical copies of this layer, forms a more accurate representation of the free fermion ground state. (b) Plot of the wavelets, together with reflections they are later paired with to form symmetric/anti-symmetric combinations (dashed), for the 2, 4 and 5 angle (or levels of unitary gates in each layer) wavelet approximations to the free fermion ground state. The 2 angle wavelets are the D4 wavelets, while the higher order wavelets look like smoothed D4 wavelets.} 
\label{fig:HigherOrder}
\end{center}
\end{figure}

\begin{figure}[!t]
\begin{center}
\includegraphics[width=8cm]{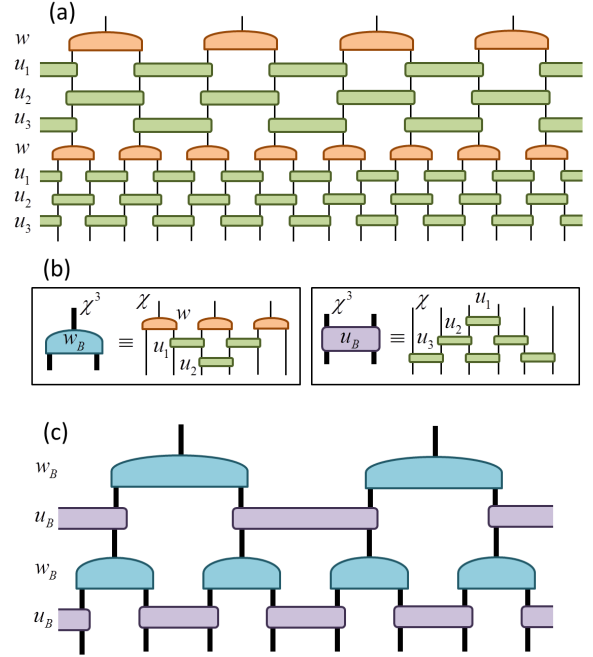}
\caption{(a) A MERA with isometries $w$ and three levels of disentanglers $\{u_1,u_2,u_3 \}$ in each layer. (b) Groups of tensors can be contracted together to form isometries $w_B$ and disentanglers $u_B$ of larger bond dimension $\chi^3$, compared to bond dimension $\chi$ of the original tensors. (c) Under the grouping from (b) the MERA with three levels of disentanglers is reduced to a standard binary MERA of larger bond dimension.} 
\label{fig:Grouping}
\end{center}
\end{figure}

In practice we find it desirable to construct custom higher-order wavelets,
rather than using standard D2N Daubechies wavelets, as custom wavelets can be designed
to allow for better low/high energy separation when later forming symmetric and
anti-symmetric wavelet pairs. We now discuss the derivation a custom wavelet
filter $q_z(r)$ of length $L=8$, which corresponds to a circuit with four
levels of unitary gates 
$\{v(\theta_1), v(\theta_2), v(\theta_3), v(\theta_4)\}$ 
in each layer, see Fig. \ref{fig:HigherOrder}(a). We find in
necessary to fix $\theta_4=-\pi/2$, such that gate $v(\theta_4)$ becomes a swap
gate, in order to later generate wavelets with a good separation of energies.
The remaining three angles $\{\theta_1, \theta_2, \theta_3 \}$ are chosen such
that the wavelets they generate have a third vanishing moment, 
\begin{equation}
{\left. {\frac{{{\partial^2 }{Q_z}}}{{\partial {k }^2}}} \right|_{k  = 0}} = 0.  \label{eq:A2}
\end{equation}
 in addition to the two vanishing moments of Eq. \ref{eq:A1}. The set of three
equations (Eqs. \ref{eq:A1} and \ref{eq:A2}) in three variables 
$\{\theta_1, \theta_2, \theta_3 \}$ can be solved algebraically, with the exact solution
given,
 \begin{align}
  \tan \left( {{\theta _1}} \right) &= \frac{{56 + 14\sqrt {106}  - \sqrt 7 \left( {23 + 2\sqrt {106} } \right)\sqrt {\left( {4\sqrt {106}  - 39} \right)} }}{{105}} \hfill \nonumber \\
  \tan \left( {{\theta _2}} \right) &= \frac{{\left( {17\sqrt 7  + 2\sqrt {742} } \right)\sqrt {\left( {4\sqrt {106}  - 39} \right)} }}{{105}} \hfill\nonumber \\
  \tan \left( {{\theta _3}} \right) &= \frac{{ - \sqrt {7\left( {4\sqrt {106}  - 39} \right)} }}{{35}} \hfill, \label{eq:A3}
\end{align}  
which yields approximate values,
\begin{equation}
\left[ {\begin{array}{*{20}{c}}
  {{\theta _1}} \\ 
  {{\theta _2}} \\ 
  {{\theta _3}} \\ 
  {{\theta _4}} 
\end{array}} \right] \approx \left[ {\begin{array}{*{20}{c}}
  {\phantom{-}0.276143653403021} \\ 
  {\phantom{-}0.950326554644286} \\ 
  { - 0.111215262156182} \\ 
  { - \pi /2} 
\end{array}} \right]. \label{eq:A4}
\end{equation}
It is not obvious that the new circuit with four levels of unitary gates in
each layer can be related to a standard binary MERA (which only has two levels
of unitary gates in each level). However, in Fig. \ref{fig:Grouping} we
demonstrate that under appropriate grouping of tensors, the circuit with four
levels of unitary gates and bond dimension $\chi$ can be precisely mapped to a
standard binary MERA with larger bond dimension $\chi^3$. [Following the same
approach a circuit with $N$ levels of unitary gates in each layer can be mapped
to a standard binary MERA of bond dimension $\chi^{(N-1)}$]. The circuit of
Fig. \ref{fig:HigherOrder}(a) for the free fermion ground state can be mapped to
a MERA of bond dimension $\chi=64$, or, following the decoupling transformation
discussed in Section C of the supplementary material, can be mapped to a
$\chi=8$ MERA for the critical Ising model. Table I of the main text the
compares the accuracy of this $\chi=8$ MERA for the Ising model, built from
higher order wavelets, to the $\chi=2$ MERA, built from the D4 wavelets. It
is seen that the $\chi=8$ MERA offers significantly improved results in all
measured quantities (including both ground energy as well as in the
reproduction of critical data for the Ising CFT).

In principle one could achieve an arbitrarily precise approximation to the free
fermion ground state by allowing more levels of unitary gates 
$\{v(\theta_1), v(\theta_2), v(\theta_3),\ldots \}$ in each layer of the circuit, and then
solving for the angles $\{\theta_1, \theta_2, \theta_3,\ldots\}$ in order to
maximize the vanishing moments of the wavelets (in addition to requiring that
the wavelets can offer a good separation of energies). In practice, however, we
have not pursued this direction much for two main reasons. (i) Firstly, beyond
four levels of unitary gates there are no closed form solutions to the
equations for vanishing moments, such that one must resort to numerical
optimization to find the appropriate angles 
$\{\theta_1, \theta_2, \theta_3,\ldots\}$. (ii) Secondly, the solutions beyond four levels of unitary
gates correspond to MERA of large bond dimension, which are not computationally
feasible to contract. This being said, we have experimented in optimizing a
five angle solution, corresponding to a $\chi=16$ MERA for the quantum critical
Ising model,
\begin{equation}
\left[ {\begin{array}{*{20}{l}}
  {{\theta _1}} \\ 
  {{\theta _2}} \\ 
  {{\theta _3}} \\ 
  {{\theta _4}} \\ 
  {{\theta _5}} 
\end{array}} \right] = \left[ {\begin{array}{*{20}{l}}
  {\phantom{-}0.133662134988773} \\ 
  { - 1.311424155804674} \\ 
  { - 0.099557657512352} \\ 
  {\phantom{-}0.717592959416643} \\ 
  {\phantom{-}0.157462489552395} 
\end{array}} \right], \label{eq:A5}
\end{equation}
where the angles were chosen to produce wavelets with good separation of
energies in addition to possessing two vanishing moments. The MERA built from
these gates reproduced a ground energy (per site) of $E = -1.272968$ for the
critical Ising model, which equates to a relative error of $0.02\%$.

\renewcommand\thefigure{C.\arabic{figure}}    
\renewcommand{\theequation}{C.\arabic{equation}}
\setcounter{figure}{0}  
\setcounter{equation}{0} 

\begin{figure}[!t]
\begin{center}
\includegraphics[width=8cm]{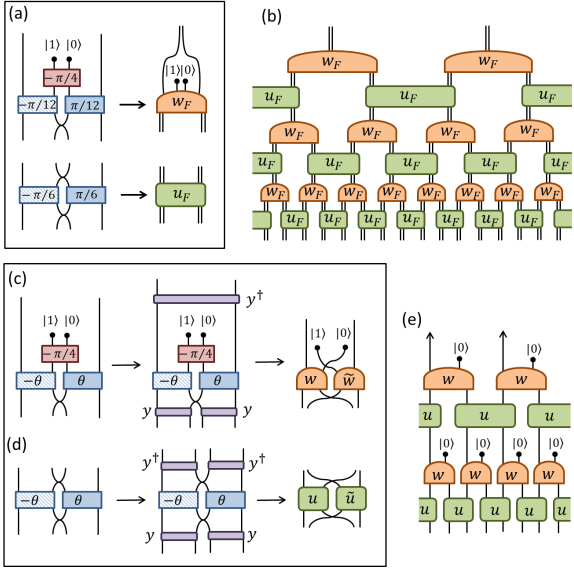}
\caption{(a) Groups of unitary gates from the circuit in Fig. \ref{fig:DoubleCircuit}(b) can be combined to form isometries $w_F$ and disentanglers $u_F$ of bond dimension $\chi=4$ (represented by a double index). (b) The isometries $w_F$ and disentanglers $u_F$ form a $\chi=4$ binary MERA for the ground state of the free fermion system in Eq. \ref{eq:B1}. (c) A unitary transformation $y$ is
made on $w_F$ from (a), where $y$ is specified as a linear mapping $Y$ of Majorana modes in Eqs. \ref{eq:B4} and \ref{eq:B5}, to decouple into a pair of tensors $w$ and $\tilde w$. (d) A unitary transformation $y$ is made on $u_F$ from (a) to decouple into a pair of tensors $u$ and $\tilde u$. (e) Tensors $w$ and $u$ form a $\chi=2$ scale-invariant binary MERA for the quantum critical Ising model.}
\label{fig:DecoupleSupp}
\end{center}
\end{figure}

\textbf{Section C: Mapping from free spinless fermions to the quantum Ising
model.---} The circuit of Fig. \ref{fig:DoubleCircuit}(b) from the main text gives an approximation to the ground state of free spinless fermions, 
\begin{equation}
{H_\textrm{ff}} = -\sum\nolimits_r {\left( {\hat a_{r + 1}^\dag {{\hat a}_r} + \hat a_r^\dag {{\hat a}_{r + 1}}} \right)}, \label{eq:B1}
\end{equation}
and can be mapped to a binary MERA of bond dimension $\chi=4$ under an appropriate grouping of tensors, see Fig. \ref{fig:DecoupleSupp}(a-b). In this section we discuss how the free (spinless) fermion Hamiltonian ${H_\textrm{ff}}$ can be decoupled into two copies of free Majorana fermions, each of which can then be mapped to the quantum critical Ising model $H_\textrm{Is.}$, see
Refs.\cite{Decouple,GroundEnt} for additional details. Furthermore, we also
describe how to implement equivalent transformations on the ground state MERA
of $H_\textrm{ff}$ in order to decouple into two copies of the
ground state MERA for the critical Ising model $H_\textrm{Is.}$.

For every spinless fermion ${\hat a}_r$ we define a pair of Majorana operators
\frw{d_{2r-1}} and \frw{d_{2r}}, 
\begin{equation}
{{\hat a}_r} = \frac{{{{\overset{\lower0.5em\hbox{$\smash{\scriptscriptstyle\smile}$}}{d} }_{2r - 1}} + i{{\overset{\lower0.5em\hbox{$\smash{\scriptscriptstyle\smile}$}}{d} }_{2r}}}}{2}, \label{eq:B2}
\end{equation}
and then express Hamiltonian ${H_\textrm{ff}}$ in these Majorana operators,
\begin{equation}
{H_\textrm{ff}} = \sum\nolimits_r {i\left( {{{\overset{\lower0.5em\hbox{$\smash{\scriptscriptstyle\smile}$}}{d} }_{2r - 1}}{{\overset{\lower0.5em\hbox{$\smash{\scriptscriptstyle\smile}$}}{d} }_{2r + 2}} - {{\overset{\lower0.5em\hbox{$\smash{\scriptscriptstyle\smile}$}}{d} }_{2r}}{{\overset{\lower0.5em\hbox{$\smash{\scriptscriptstyle\smile}$}}{d} }_{2r + 1}}} \right)}. \label{eq:B3}
\end{equation}
The Majorana representation of ${H_\textrm{ff}}$ can then be decoupled by
unitarily transforming blocks of four Majorana modes,
\begin{equation}
{{\overset{\lower0.5em\hbox{$\smash{\scriptscriptstyle\smile}$}}{d'}}_{2r - 1 + k}} = \sum\limits_{p = 1,2,3,4} {{Y_{kp}}{{\overset{\lower0.5em\hbox{$\smash{\scriptscriptstyle\smile}$}}{d} }_{2r - 1 + p}}}, \label{eq:B4}
\end{equation}
where matrix $Y$ is defined,
\begin{equation}
Y = \left[ {\begin{array}{*{20}{c}}
  0&1&0&0 \\ 
  0&0&1&0 \\ 
  1&0&0&0 \\ 
  0&0&0&1 
\end{array}} \right]. \label{eq:B5}
\end{equation}
The transformed Hamiltonian $H'_\textrm{ff}$, written in terms of Majorana
modes \frw{{d'}_{2r}}, has been decoupled into two copies of free Majorana
fermions, ${{\tilde H}_\textrm{ff}} = {H_A} + {H_B}$, with
\begin{equation}
{H_A} = \sum\limits_{r\textrm{ odd}} {i\left( {{{\overset{\lower0.5em\hbox{$\smash{\scriptscriptstyle\smile}$}}{d'} }_{2r}}{{\overset{\lower0.5em\hbox{$\smash{\scriptscriptstyle\smile}$}}{d'} }_{2r + 3}} - {{\overset{\lower0.5em\hbox{$\smash{\scriptscriptstyle\smile}$}}{d'} }_{2r - 1}}{{\overset{\lower0.5em\hbox{$\smash{\scriptscriptstyle\smile}$}}{d'} }_{2r}}} \right)}, \label{eq:B6}
\end{equation}
supported on the $r$ odd sublattice, and ${H_B}$ defined identically but
supported on the $r$ even sublattice (see Ref.\cite{Decouple} for additional
details). Each of the Hamiltonians $H_A$ and $H_B$ can then be mapped to a copy
of the quantum critical Ising model $H_\textrm{Is.}$ via the Jordan-Wigner
transformation \cite{GroundEnt}.

The decoupling transformation can similarly be implemented on the quantum
circuit, see Fig. \ref{fig:DoubleCircuit}(b), that approximated the ground state
of $H_\textrm{ff}$. The collection of gates considered in
Fig. \ref{fig:DecoupleSupp}(c), which could be contracted to form an
isometry $w_F$ of the free fermion MERA as depicted in Fig. \ref{fig:DecoupleSupp}(a), can be represented as a linear map on four
spinless fermion modes  $\hat a$, 
\begin{equation}
{{\hat a'}_r} = \sum\limits_{p = 1,2,3,4} {{{\left[ {{W_F}(\theta )} \right]}_{rp}}{{\hat a}_p}}, 
\end{equation}
where matrix ${W_F}(\theta )$ is defined,
\begin{equation}
{W_F}(\theta ) = \frac{1}{{\sqrt 2 }}\left[ {\begin{array}{*{20}{c}}
  {\sqrt 2 {c_\theta }}&{{s_\theta }}&{ - {s_\theta }}&0 \\ 
  0&{{c_\theta }}&{{c_\theta }}&{\sqrt 2 {s_\theta }} \\ 
  {\sqrt 2 {s_\theta }}&{ - {c_\theta }}&{{c_\theta }}&0 \\ 
  0&{ - {s_\theta }}&{ - {s_\theta }}&{\sqrt 2 {c_\theta }} 
\end{array}} \right], \label{eq:B7}
\end{equation}
with $c_\theta $ and  $s_\theta $ short for $\cos(\theta)$ and $\sin(\theta)$ respectively. 
Similarly, the collection of gates considered in Fig. \ref{fig:DecoupleSupp}(d), 
which can be contracted to form a disentangler $u_F$ of the free fermion MERA as shown in Fig. \ref{fig:DecoupleSupp}(a), 
is also a linear map on four fermionic modes $\hat a$, here defined by the matrix,
\begin{equation}
{U_F}(\theta ) = \left[ {\begin{array}{*{20}{c}}
  {{c_\theta }}&0&{ - {s_\theta }}&0 \\ 
  0&{{c_\theta }}&0&{{s_\theta }} \\ 
  {{s_\theta }}&0&{{c_\theta }}&0 \\ 
  0&{ - {s_\theta }}&0&{{c_\theta }} 
\end{array}} \right]. \label{eq:B8}
\end{equation}
Each of these linear maps, defined by ${W_F}(\theta )$ and ${U_F}(\theta )$, can be 
decoupled into a pairs of separate maps by first (i) re-writing the maps in terms of 
Majorana modes, then (ii) applying the decoupling transformation $Y$, described by 
Eqs. \ref{eq:B4} and \ref{eq:B5}, as depicted in Fig. \ref{fig:DecoupleSupp}(c,d). 
The result is that ${W_F}(\theta )$, which was a map for four spinless fermion modes 
(or eight Majorana modes), is decoupled into a pair of maps ${W_M}\left( \theta  \right)$ 
each acting on four Majorana modes,
\begin{equation}
{{\overset{\lower0.5em\hbox{$\smash{\scriptscriptstyle\smile}$}}{d'} }_{2r - 1 + k}} = \sum\limits_{p = 1,2,3,4} {{{\left[ {{W_M}(\theta )} \right]}_{rp}}{{\overset{\lower0.5em\hbox{$\smash{\scriptscriptstyle\smile}$}}{d} }_{2r - 1 + p}}}, \label{eq:B9}
\end{equation}
where matrix ${W_M}\left( \theta  \right)$ is defined,
\begin{equation}
{W_M}\left( \theta  \right) = \left[ {\begin{array}{*{20}{c}}
  {{c_\theta }}&0&{ - {s_\theta }}&0 \\ 
  0&{{s_\theta }}&0&{ - {c_\theta }} \\ 
  {{s_\theta }}&0&{{c_\theta }}&0 \\ 
  0&{{c_\theta }}&0&{{s_\theta }} 
\end{array}} \right]. \label{eq:B10}
\end{equation}
Likewise, ${U_F}(\theta )$ is decoupled into a pair of maps ${U_M}\left( \theta  \right)$, 
each acting on four Majorana modes,
\begin{equation}
{U_M}(\theta ) = \left[ {\begin{array}{*{20}{c}}
  {{c_\theta }}&0&{ - {s_\theta }}&0 \\ 
  0&{{c_\theta }}&0&{{s_\theta }} \\ 
  {{s_\theta }}&0&{{c_\theta }}&0 \\ 
  0&{ - {s_\theta }}&0&{{c_\theta }} 
\end{array}} \right]. \label{eq:B11}
\end{equation}
The linear maps ${W_M}(\theta )$ and ${U_M}(\theta )$ for 
Majorana modes can be re-written, 
via the Jordan-Wigner transform, as 
unitaries $w(\theta )$ and $u(\theta )$ acting on two spin-$1/2$ sites 
(since four Majorana modes equals two spins). This yields,
\begin{align}
w(\theta ) & = \left[ {\begin{array}{*{20}{c}}
  {{c_{(\theta  - \pi /4)}}}&0&0&{{s_{(\theta  - \pi /4)}}} \\ 
  0&{{c_{(\pi /4)}}}&{ - {s_{(\pi /4)}}}&0 \\ 
  0&{{s_{(\pi /4)}}}&{{c_{(\pi /4)}}}&0 \\ 
  { - {s_{(\theta  - \pi /4)}}}&0&0&{{c_{(\theta  - \pi /4)}}} 
\end{array}} \right], \label{eq:B12}
\end{align}
\begin{align}
u(\theta ) & = \left[ {\begin{array}{*{20}{c}}
  {{c_\theta }}&0&0&{{s_\theta }} \\ 
  0&1&0&0 \\ 
  0&0&1&0 \\ 
  { - {s_\theta }}&0&0&{{c_\theta }} 
\end{array}} \right], \label{eq:B13}
\end{align}
when written in the basis of Pauli $Z$ eigenstates, 
$\left\{ {\left| { \uparrow  \uparrow } \right\rangle ,\left| { \downarrow  \uparrow } \right\rangle ,\left| { \uparrow  \downarrow } \right\rangle ,\left| { \downarrow  \downarrow } \right\rangle } \right\}$. 
These tensors can alternatively be expressed as sums of tensor products of Pauli matrices,
\begin{align}
w(\theta) = & \frac{{\sqrt 2 {c_{(\theta  - \pi /4)}} + 1}}{{2\sqrt 2 }}II + \frac{{\sqrt 2 {c_{(\theta  - \pi /4)}} - 1}}{{2\sqrt 2 }}ZZ + \nonumber \\
& i\frac{{1 - \sqrt 2 {s_{(\theta  - \pi /4)}}}}{{2\sqrt 2 }}XY - i\frac{{1 + \sqrt 2 {s_{(\theta  - \pi /4)}}}}{{2\sqrt 2 }}YX \nonumber \\
u(\theta) = & \frac{{{c_\theta } + 1}}{2}II + \frac{{{c_\theta } - 1}}{2}ZZ - \frac{{i{s_\theta }}}{2}XY - \frac{{i{s_\theta }}}{2}YX, \label{eq:B14}
\end{align}
where $ZZ$ is short for $Z\otimes Z$ etc. 
The $\chi=2$ MERA of Eq. \ref{eq:16} in the main text corresponds to setting 
isometries $w(\theta_1)$ and disentanglers $u(\theta_2)$ with angles 
$[\theta_1,\theta_2] = [\pi/12,-\pi/6]$ 
associated to the D4 Daubechies wavelets. 
Similarly, the $\chi=8$ MERA considered in Table I of the main text, 
corresponds to constructing unitary gates $\{w(\theta_1), u(\theta_2), u(\theta_3), u(\theta_4)\}$, 
with angles $\theta$ as specified in Eq. \ref{eq:A4}, and then grouping sets of tensors to form a 
MERA of bond dimension $\chi=8$ as described in Fig. \ref{fig:Grouping}. 

The tensors derived in Eq. \ref{eq:B14} represent the scale-invariant part of a
MERA that approximates the ground state of the quantum critical Ising model.
However, one also needs to consider the transitional layer, i.e $V$ of
Fig.  \ref{fig:DoubleCircuit}(b), in order to build the full ground-state MERA.
When transformed into spin degrees of freedom the transition layer $V$ becomes
a product of commuting gates $v_{r,r + 1}$,
\begin{equation}
V = \prod\limits_r {{v_{r,r + 1}}},
\end{equation}
which are defined,
\begin{equation}
{v_{r,r + 1}} = \tfrac{1}{{\sqrt 2 }}\left( { - {I_r}{I_{r + 1}} + i{X_r}{X_{r + 1}}} \right). \label{eq:B15}
\end{equation}
Together, the transition layer of tensors $V$ in addition the scale-invariant
MERA built from gates $w(\theta)$ and $u(\theta)$ constitute the approximation
to the ground state of the Ising model,
\begin{equation}
{H_\textrm{Is.}} = \sum\limits_r {\left( { - {Z_r} - {X_r}{X_{r + 1}}} \right)}. \label{eq:B16}
\end{equation}
Alternatively, we can transform the Ising model, 
${\tilde{H}_\textrm{Is.}} \equiv {V^\dag }{H_\textrm{Is.} }V$, where the transformed Hamiltonian
${\tilde{H}_\textrm{Is.}}$ is given as,
\begin{equation}
{\tilde H}_\textrm{Is.} = \sum\limits_r {\left( {{X_r}{Z_{r + 1}}{X_{r + 2}} - {X_r}{X_{r + 1}}} \right)}, \label{eq:B17}
\end{equation}
such that the ground state of ${\tilde{H}_\textrm{Is.}}$ 
can be approximated as a fully scale-invariant MERA (i.e. with no transitional layers).

\renewcommand\thefigure{D.\arabic{figure}}    
\renewcommand{\theequation}{D.\arabic{equation}}
\setcounter{figure}{0}  
\setcounter{equation}{0} 

\textbf{Section D: RG flow of the Hamiltonian.---}
In this section we discuss in more detail the RG flow of the critical Ising Hamiltonian generated from the wavelet based MERA. For simplicity we focus on the $\chi = 2$ MERA of Eq. \ref{eq:16}, although similar results also follow for the $\chi=8$ MERA discussed in Section B. 

\begin{figure}[!t]
\begin{center}
\includegraphics[width=8cm]{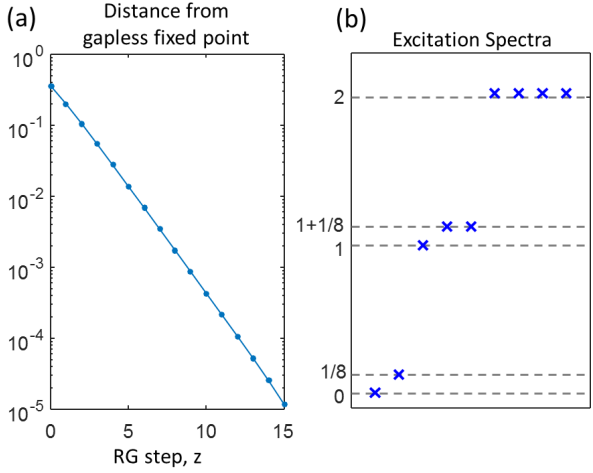}
\caption{(a) Distance of the Hamiltonian coupling $h^{[z]}$ at scale $z$ from the fixed point coupling $h^{*}$, see Eq. \ref{eq:D5}. (b) The excitation spectrum (above the ground state) of the fixed point Hamiltonian $H_\textrm{Is.}^{*}$, which is seen to closely approximate the excitation spectrum of the quantum critical Ising model in the thermodynamic limit.  } 
\label{fig:HamCG}
\end{center}
\end{figure}

Let $\tilde H_\textrm{Is.}^{[0]}=\sum\nolimits_r h_{(r,r+1,r+2)}^{[0]}$ be the critical Ising Hamiltonian after transformation by the transitional layer of the MERA, with three body coupling defined,
\begin{equation}
h_{(r,r+1,r+2)}^{[0]} \equiv \sum\limits_r {\left( {{X_r}{Z_{r + 1}}{X_{r + 2}} - {X_r}{X_{r + 1}}} \right)}, \label{eq:D0}
\end{equation}
see also Eq. \ref{eq:B17} in section C. We normalize $\tilde H_\textrm{Is.}^{[0]}$ by adding an appropriate contribution of the identity such that
\begin{equation}
\bra{\psi} \tilde H_\textrm{Is.}^{[0]} \ket{\psi} = 0, \label{eq:D1}
\end{equation}
with $\ket{\psi}$ the scale-invariant part of the MERA (i.e. the scale-invariant MERA defined by the isometries $w$ and disentanglers $u$ of Eq. \ref{eq:16} without any transitional layers). Each layer of the scale-invariant MERA can be regarded as a step of coarse-graining with entanglement renormalization, which can be applied to generate a sequence of coarser Hamiltonians,  
\begin{equation}
\tilde H_\textrm{Is.}^{[0]} \to \tilde H_\textrm{Is.}^{[1]} \to \tilde H_\textrm{Is.}^{[2]} \to \tilde H_\textrm{Is.}^{[3]} \to \ldots. \label{eq:D2}
\end{equation}
Here the Hamiltonian at scale $z$ is of the form $\tilde H_\textrm{Is.}^{[z]}=\sum\nolimits_r h_{(r,r+1,r+2)}^{[z]}$, where the local couplings $h^{[z]}$ are defined recursively, 
\begin{equation}
h^{[z+1]} = 4\bar{\mathcal{S}} \left(  h^{[z]} \right), \label{eq:D3}
\end{equation}
with $\bar{\mathcal{S}}$ the average scaling superoperator associated to the scale-invariant MERA \cite{MERAbook}. Note that the factor of 4 in Eq. \ref{eq:D3} allows the couplings $h^{[z]}$ to have a consistent norm over different scales by offsetting the scaling dimension, $\Delta = 2$, of the Hamiltonian. The fixed point Hamiltonian $\tilde H_\textrm{Is.}^{*}=\sum\nolimits_r h_{(r,r+1,r+2)}^{*}$ is defined in terms of coupling
\begin{equation}
{h^*} \equiv \mathop {\lim }\limits_{z \to \infty } \left( {{h^{[z]}}} \right), \label{eq:D4}
\end{equation}
which can be evaluated directly by decomposing $h^{[0]}$ in the basis of eigenoperators of $\bar{\mathcal{S}}$ and then discarding components corresponding to sub-leading eigenvalues. Here it is found that $h^{[0]}$ has strictly zero component for all scaling operators of smaller scaling dimension than that of the Hamiltonian (i.e. that it contains no RG relevant terms that would shift the flow from criticality).

Let us define the distance $D$ from the fixed point as, 
\begin{equation}
D(z) \equiv \norm{h^{[z]} - h^{[*]}}/ \norm{h^{[*]}}. \label{eq:D5}
\end{equation}
Fig. \ref{fig:HamCG}(a), which displays $D$ as a function of RG step, shows that the Hamiltonian flows smoothly to the fixed point Hamiltonian $\tilde H_\textrm{Is.}^{*}$ under coarse-graining. The excitation spectra of $\tilde H_\textrm{Is.}^{*}$, when diagonalized on a lattice of 10 sites and scaled such that the smallest excitation has value $1/8$, is plotted in Fig. \ref{fig:HamCG}(b). Here it is seen that $\tilde H_\textrm{Is.}^{*}$ possesses, to good approximation, the correct excitation spectrum of the quantum critical Ising model in the thermodynamic limit \cite{CFTbook1,CFTbook2}. 
These results demonstrate that, under coarse-graining with the wavelet based MERA, the Hamiltonian of the quantum critical Ising model flows to a gapless fixed point that captures the correct low energy behavior of the Ising model. That the wavelet based MERA generates a flow to a gapless fixed point follows from the restriction that the wavelets precisely vanish at the Fermi points, see Eq. \ref{eq:10}, which prevents the introduction of RG relevant terms into the effective Hamiltonians.


\begin{thebibliography}{99}
\bibitem{WaveNorm}
We are assuming the orthogonal wavelets are normalized to unity (under scalar
product with themselves) such that they form a unitary change of basis.
\end{thebibliography}
\end{document}